\documentclass[review,authoryear,12pt]{elsarticle}

 \usepackage{graphicx}
 \usepackage{verbatim} 
 \usepackage{enumitem,pifont,xcolor}
 \usepackage{lineno}

\graphicspath{ {enflfigures/} }

\usepackage[nolists]{endfloat}
\usepackage{lipsum}

\usepackage{changes}
\colorlet{Changes@Color}{red}

\AtBeginDelayedFloats{\linespread{2}} 

\usepackage{amssymb}
\usepackage{amsmath} 

  \usepackage{geometry}
   \usepackage{float}
    \usepackage{booktabs}

\journal{Journal of Waterway, Port, Coastal, and Ocean Engineering}

\begin{document}
\begin{frontmatter}

\title{\textbf{The Temporal and Spatial Evolution of Momentum, Kinetic Energy and Force in Tsunami Waves during Breaking and Inundation}}

\author[1]{Roberto Marivela}
\author[2]{Robert Weiss}
\author[3]{Costas E. Synolakis}

\address[1]{Ph.D. Candidate, Department of Geosciences, Virginia Tech. \\ 4044 Derring Hall (0420)
Blacksburg, VA 24061, USA \\ {robertom@vt.edu} \\ }

\address[2]{Associate Professor, Department of Geosciences, Virginia Tech. \\ 4044 Derring Hall (0420)
Blacksburg, VA 24061, USA \\ {weiszr@vt.edu}}

\address[3]{Professor, Viterbi School of Engineering, University of Southern California. \\3620 S. Vermont Avenue, KAP 206C Los Angeles, CA 90089-2531, USA \\costas@usc.edu }

\begin{abstract}

A plethora of studies in the past decade describe tsunami hazards and study their evolution from the source to the target coastline, but mainly focus on coastal inundation and  maximum runup. Nonetheless, anecdotal reports from eyewitnesses, photographs and videos suggest counterintuitive flow dynamics, for example rapid initial acceleration when the wave first strikes the initial shoreline. Further, the details of the flow field at or within tens of meters of the shoreline are exquisitely important in determining damage to structures and evacuation times. Based on a set of three-dimensional numerical simulations using solitary waves as a model, we show  the spatial-temporal distribution of the flow momentum, kinetic energy and force during the breaking process. We infer that the flow reaches its highest destructive capacity not when flow momentum or kinetic energy reach their maxima, but when flow force reaches its. This occurs in the initial shoreline environment, which needs to be considered in nearshore structures design.

\end{abstract}


\end{frontmatter}

					\section*{Introduction} 

The 2004 Sumatra tsunami struck the Indian Ocean and presented a wake-up call around the globe for improved preparedness and awareness for tsunamis \citep{synolakis2006tsunami} and other coastal hazards. In its aftermath, many countries initiated the assessment or, in some cases reassessment, of their tsunami risks which mainly have been based on the maximum runup determination. \cite{synolakis1987runup} has been broadly referenced for this purpose, who developed an analytical model for non-breaking waves runup. However, the approach for breaking analysis must be numerical due to its complexity.

In their report for U.S. coastlines, \citet{dunbar2008us} mentioned the importance of numerical modeling in hazard assessments to quantify the impacts of future events. It has been recognized that as long as waves are not breaking, Boussinesq and shallow water equations can be applied to simulate tsunami wave dynamics. Especially the Boussinesq equations are appropriate to study the wave approach and the runup process, and significant efforts have been made to include the proper dispersive terms \citep{madsen2006boussinesq, roeber2012boussinesq, lynett2002modeling, lynett2004two}. For a better representation of the flow field characteristics with Boussinesq equations, \citet{lynett2004linear} proposed to use multiple layers along the vertical direction. However, such a multi-layer high-order Boussinesq model is computationally expensive. Furthermore, very close to the coast where the waves break, the irrotational assumption, which is appropriate for non-breaking waves, is violated. Therefore, to fully understand and comprehensively study the very near-shore dynamics and effects of tsunami waves is better to explore with three-dimensional computational models.

Additionally, \citet{dunbar2008us} also highlighted that the runup value alone might not be appropriate to quantify the damage caused by tsunamis. Runup describes the inundated area to the first order and can be employed for a large-scale overview of what happens after a tsunami strikes. However, runup values might not be sufficient to explore better coastal management strategies and solutions.

In this contribution, we use Lagrangian numerical simulations to revisit the wave breaking hydrodynamics. We simulate the experimental setup for the canonical problem for long-wave runup with GPUSPH. We utilize three-dimensional solitary waves in order to be consistent with \cite{synolakis1986runup,synolakis1987runup} even though \citet{madsen2008solitary} and \citet{,madsen2010analytical} discussed the fact that solitary waves are not the best model for tsunami waves.

				\section*{Theoretical background} 

        \subsection*{The GPUSPH model} 

We use GPUSPH, a computer code that employs Smoothed Particle Hydrodynamics (SPH) to simulate breaking and non breaking solitary waves. SPH solves the Navier-Stokes equations aided by the computational resource of graphical processing units \citep[GPU,][]{herault2010sph,GPUSPH}.
Because of its Lagrangian nature, SPH is an appropriate approach to simulate flows with high turbulence such as breaking waves \citep{dalrymple2006numerical}. Based on the general SPH formulation, the motion equations are written as \citet{monaghan1992smoothed}:

\begin{equation}\label{SPHdensity}
  \begin{aligned}
\frac{\ d\rho_a}{dt}= \sum_{b}m_b(v_a-v_b)\cdot\nabla W_{ab} \\
\frac{\ dv_a}{dt}=-\sum_{b}m_b\Bigg(\frac{p_b}{\rho_b^2}+\frac{\rho_a}{\rho_a^2}+\Pi_{ab}\Bigg)\nabla W_{ab}
 \end{aligned}
\end{equation}
in which $\rho$, $m$, $v$ and $p$ represent the density, mass, velocity and pressure of the fluid particle $a$ and its neighboring fluid particles $b$. The distance between particles $a$ and $b$ is represented by $r_{ab}$, $W_{ab}$ is the kernel or interpolation function and $\Pi_{ab}$ is the artificial viscosity to prevent spurious particle movements \citep{monaghan1992smoothed}:

\begin{equation} 
 \Pi_{ab} =
  \begin{cases}\label{Pi}
   \dfrac {-\alpha c_{ab} \mu_{ab}+\beta \mu_{ab}^2 }{\rho_{ab}} & \text{if } v_{av}\cdot r_{ab} < 0\\
   0       & \text{if } v_{av}\cdot r_{ab} > 0
  \end{cases}
\end{equation}
the parameter $c$ is the speed of sound. For surface flows, the parameters $\alpha$ and $\beta$ are constants with values of 0.01 and 0 respectively \citep{monaghan1994simulating}. The initial distance between particles, $h$, is constant. For each particle to consider its neighbors, the radius of the support domain is $2-h$, and therefore related to the initial particle distribution. All particles within the kernel are referred to as neighbor particles and form the neighbor list for the calculation of the physical properties. Due to the relative movement of particles, neighbor lists need to be updated at very time step. Due to the fact that Eq. (\ref{SPHdensity}) is weakly compressible, an equation of state is required to relate pressure to density \citep{monaghan1992smoothed}:

\begin{equation}\label{SPHpressure}
p_i=\frac{\rho_0\cdot c^2}{\gamma}\Bigg(\Bigg(\frac{p_i}{p_0}\Bigg)^\gamma-1\Bigg)
\end{equation}
For a more complete description of the SPH method, we refer to \citet{gingold1977smoothed}, \citet{monaghan1992smoothed} and \citet{monaghan1994simulating}.

        \subsection*{The canonical problem for long-wave runup} 
        
\begin{figure}[h] 
\centering
\includegraphics[width=1.0\textwidth]{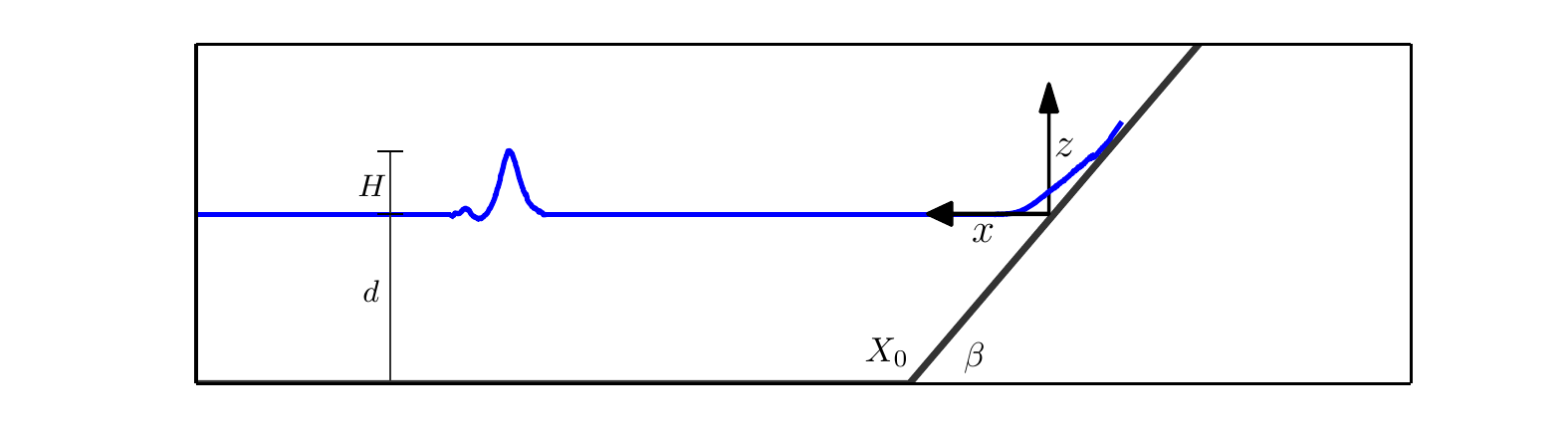} 
\caption{Domain sketch of the experiments adapted from \cite{synolakis1986runup} and \cite{synolakis1987runup} and used for our GPUSPH simulations. The toe of the beach is represented by $X_0$ and located at 14.68 m  from the initial wave maker position. $H$ is the wave height measured from the initial depth, $d$. The beach slope is $\beta=1:19.85$. Coordinate system origin is located at the initial beach shoreline.}
\label{Scheme}
\end{figure}

        We employ the experimental setup of \citet{synolakis1986runup,synolakis1987runup} for the numerical simulations with GPUSPH (Fig.~\ref{Scheme}). The particle diameter is $0.010$ m. The tank is 37.73 m long, 0.39 m wide, and 0.61 m high. The toe of the beach, $X_0$, is located at 14.68 m measured from the initial wave maker location. Parameter $\beta$  represents the slope of the beach ($\beta=1:19.85$). $H$ is the wave height measured from the initial depth, $d$. The origin of the coordinate system is located at the initial position of the shoreline, $x$ increases towards the wave maker and $z$ increases upward. The solitary waves have the following surface profile \citep{synolakis1987runup}:

\begin{equation}
\label{mu}
\eta(x,0)=\frac{H}{d}\,sech^2\bigg[\sqrt{3H/4d}(x-X_1)\bigg]
\end{equation}
where $\eta$ is the wave height and $X_1$ refers to the location where the wave elevation corresponds to the amplitude of the solitary  wave, $H$. From Eq. \ref{mu}, \cite{synolakis1990generation} obtained the following wave maker formula:

\begin{equation}\label{WMdisplacement}
\frac{d\delta}{dt}=\frac{Hsech^2\,k(\delta-ct)}{1+Hsech^2\,k(\delta-ct)}
\end{equation}
in which $\delta$ is the wave maker displacement and $k$ is the wave number ($k=\sqrt{3H/4d^3}$).

The piston is represented by boundary particles regularly distributed and applies Lennard-Jones boundary condition \citep{jones1924determination,monaghan1994simulating}. The beach, the bottom surfaces and the domain sides are represented by the method proposed in \citet{monaghan2009sph} which applies a smoothing kernel to the boundaries.

		\subsection*{Data analysis of the GPUSPH simulations} 

	To analyze the results of the simulations, we divide the flume domain into segments. 
Each segment is $0.01$ m thick ($x$ direction in Fig. 1), 0.39 m wide and 0.70 m high (width and height of the flume respectively). The analysis time increment, $\delta t$, is 0.10 s. For a certain $\delta t$, we compute in each segment the averaged flow momentum, averaged flow kinetic energy and averaged flow force \citep[hydrodynamic force]{fema2011} per unit volume of fluid contained in the segment. In computing the flow forces, the drag coefficient, $C_d$, is assumed to be 1 \citep[Table 8-2]{fema2011} as we are not considering any physical objects in the flow domain. Then we pick the maximum values and their locations of the aforementioned variables. We refer to these maximum values as the maximum flow momentum, maximum flow kinetic energy and maximum flow force. We repeat this process at each $\delta t$ so we track the maxima in space and time. Then we obtain the absolute maximum of the flow momentum, flow kinetic energy and flow force from all $\delta t$ maxima. For data analysis, we employ dimensionless time, $t^*=t\sqrt{g/d}$, dimensionless length, $x^*=x/d$, and dimensionless wave height,	 $\eta^*=\eta/d$.

	Additionally, we use the wave crest, $W_c$, defined by the water elevation that represents the wave amplitude at each $\delta t$. The tracking of the wave crest during the simulation defines the wave crest path. Wave front, $W_f$, is defined by the wave bore whose elevation is 40$\%$ of $H$. The 40$\%$ threshold is obtained by comparing the wave front of case $H/d=0.0185$ shown in Fig. 3.5.5 of Synolakis (1986) and the SPH simulation with the same setup. The best fit is determined by defining wave front as 40$\%$ of $H$. We keep this criterion for the rest of $H/d$ cases studied in this work.

		\section*{Results}	

		\subsection*{Validation of GPUSPH}	
			
\begin{figure} 
\centering
\includegraphics[width=1.0\textwidth]{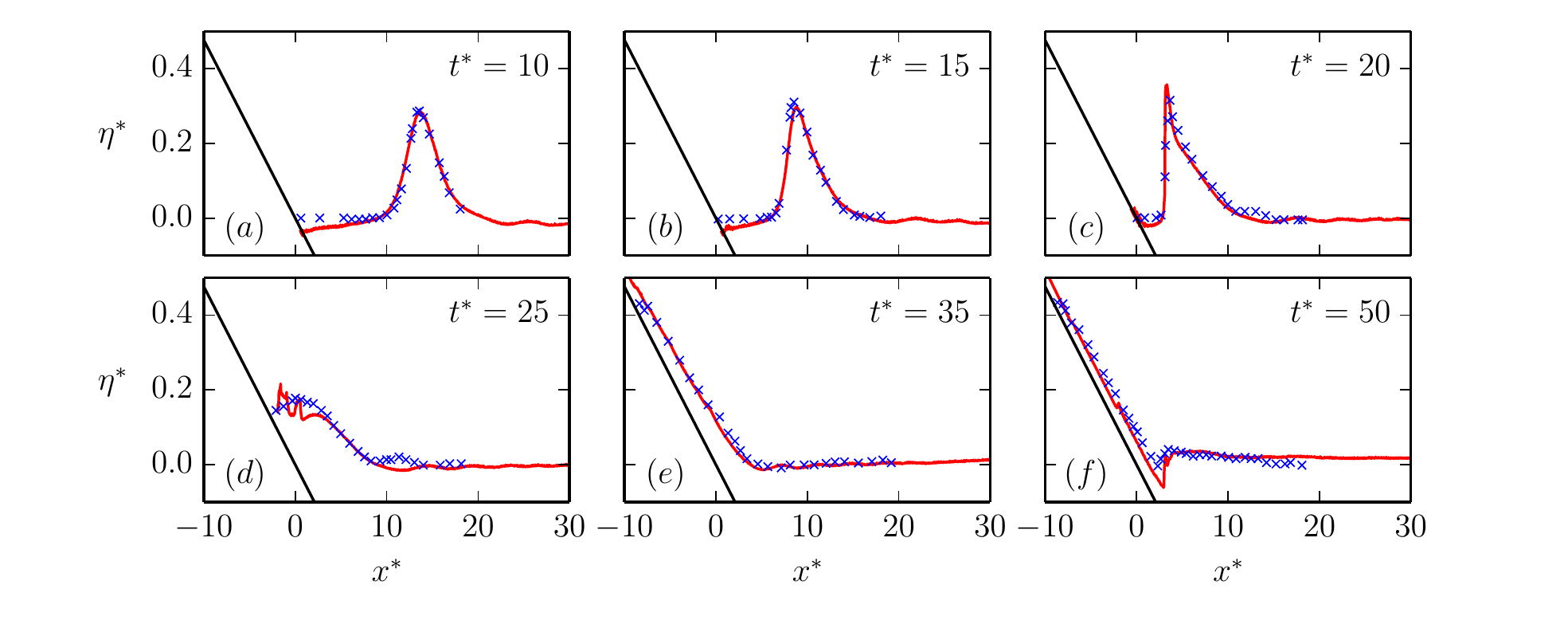} 
\caption{Comparison of surface profiles of $H/d=0.30$ solitary waves up 1:19.85 beach between the 225a experiment adapted from \cite{synolakis1986runup} (crosses) and the numerical simulation (red solid line).}
\label{Validation}
\end{figure}

\begin{figure} 
\centering
\includegraphics[width=0.7\textwidth]{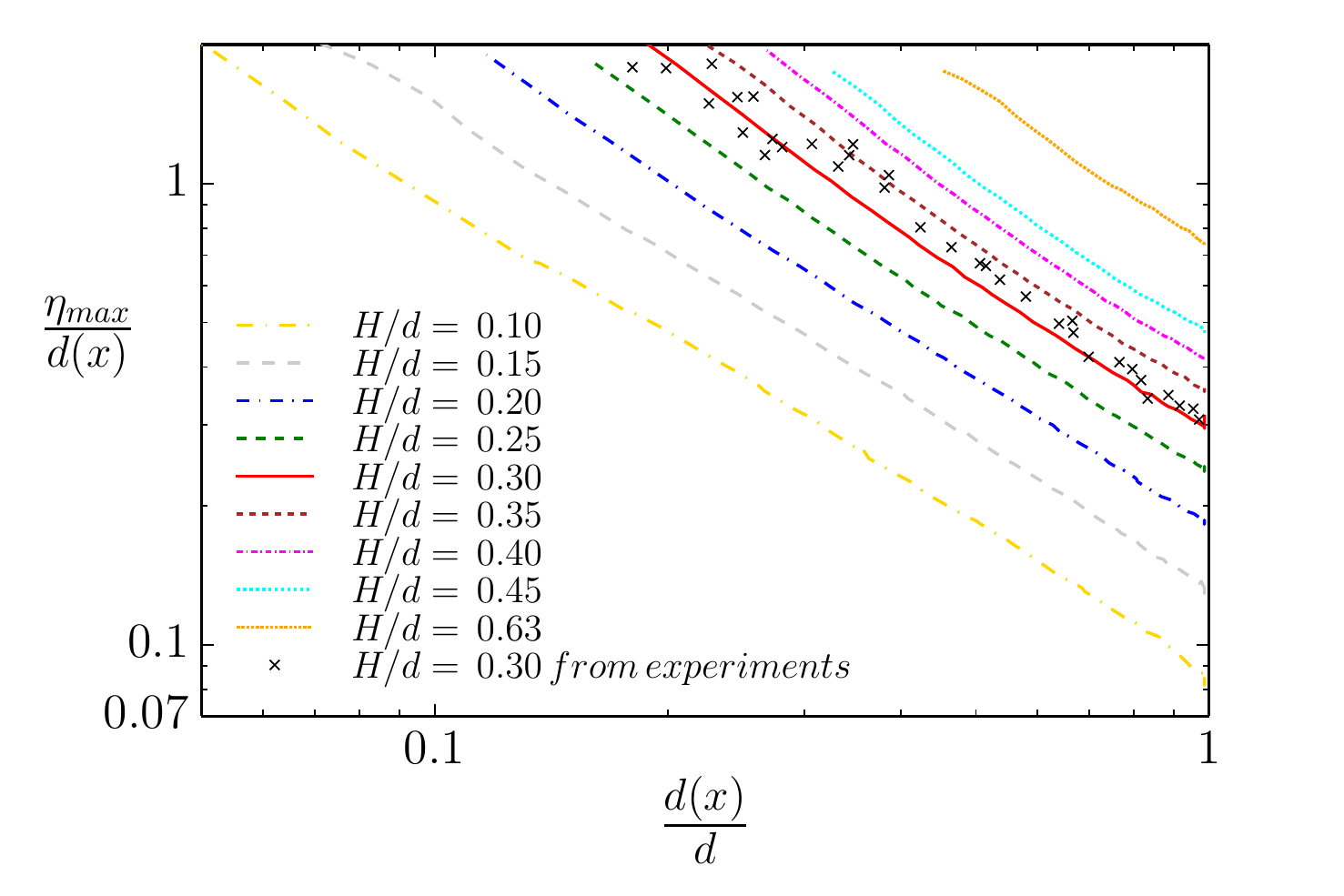} 
\caption{Maximum amplitude of different solitary waves simulations (lines) and from the experiment $H/d =0.3$ adapted from \cite{synolakis1993evolution} up 1:19.83 beach.}
\label{MaximumHeight}
\end{figure}

To validate GPUSPH for the canonical problem, we simulate the case 225a from \cite{synolakis1986runup}. The case 225a refers to experiments with a depth of $d=0.1962$ m and $H/d=0.30$. This is a case with strong breaking. In Fig.~\ref{Validation} we compare  the experiment 225a (crosses) and GPUSPH simulation (red lines) before, during and after wave breaking. The fit between the measurements from the laboratory experiment and simulation is excellent, even after breaking. Furthermore, in Fig.~\ref{MaximumHeight} we compare the distribution of the maximum wave heights between the solitary wave $H/d=0.30$ from \citet[black crosses]{synolakis1993evolution} and our simulation (solid line). The good fit between the laboratory experiments and the numerical simulation appears. We also present the distribution of the maximum wave amplitudes for the rest of the simulations developed in this work.

A further validation is carried out by comparing the velocities distribution between the experiments presented in Irish et al. (2014) and the SPH simulations. The setup is 1:10 steep beach whose toe is located at the end of a horizontal plane, 22 m long from the initial wave maker location ($x$ = 0). The initial basin depth is 0.73 m measured from the horizontal plane and the wave height is 0.43 m. Velocities from experiments are measured using acoustic Doppler velocimeters (ADVs; 50 Hz) at locations $x$ = 32.87 m (A) and $x$ = 35.06 m (B) respectively. Figure 4 shows the horizontal velocity comparison between the Irish et al. (2014) experiments and our GPUSPH simulations. This good fit also indicates the suitability of the SPH particle size to simulate the solitary waves presented in this work. Notice that there is a lack of data from experiments during initial runup compared with SPH. This is because the bubbly and turbulent bore resulted in noisy measurements from ADVs.

\begin{figure}  
\centering
\includegraphics[width=0.6\textwidth]{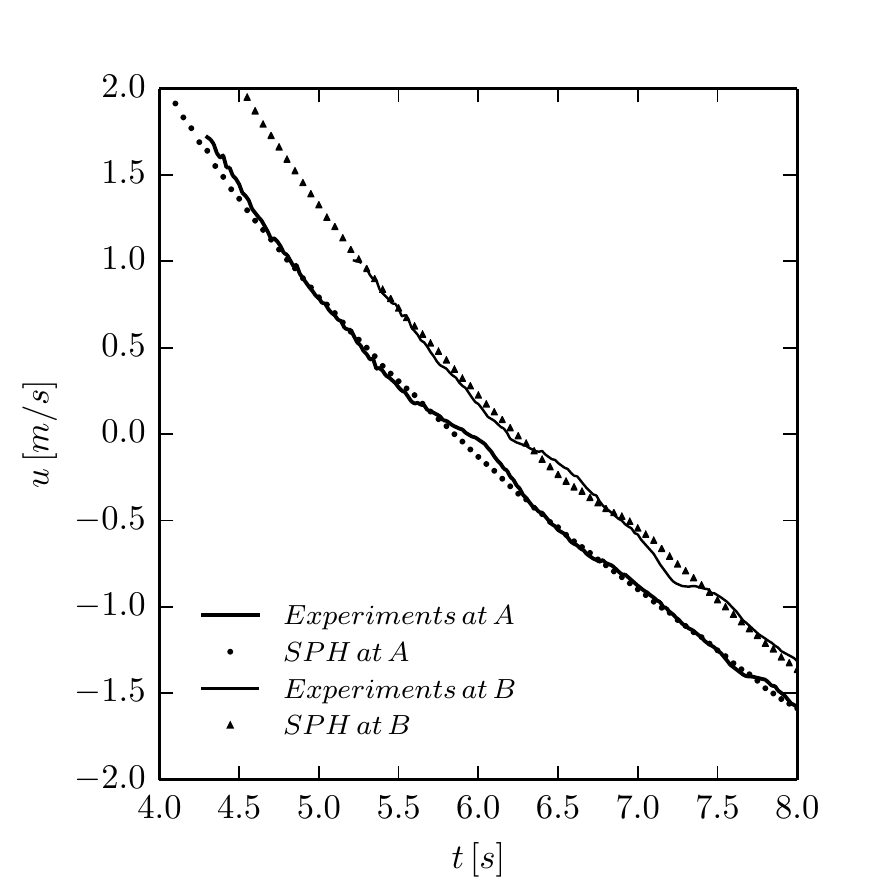} 
\caption{$u$-velocity comparison between the experiments \citep{irish2014laboratory} and the SPH simulations at locations $x$ = 32.87 m (A) and $x$ = 35.06 m (B) respectively from the initial wave maker location ($x$ = 0).}
\label{VelCom}
\end{figure}

\subsection*{Solitary waves results}	

\begin{figure} 
\centering
\includegraphics[width=0.7\textwidth]{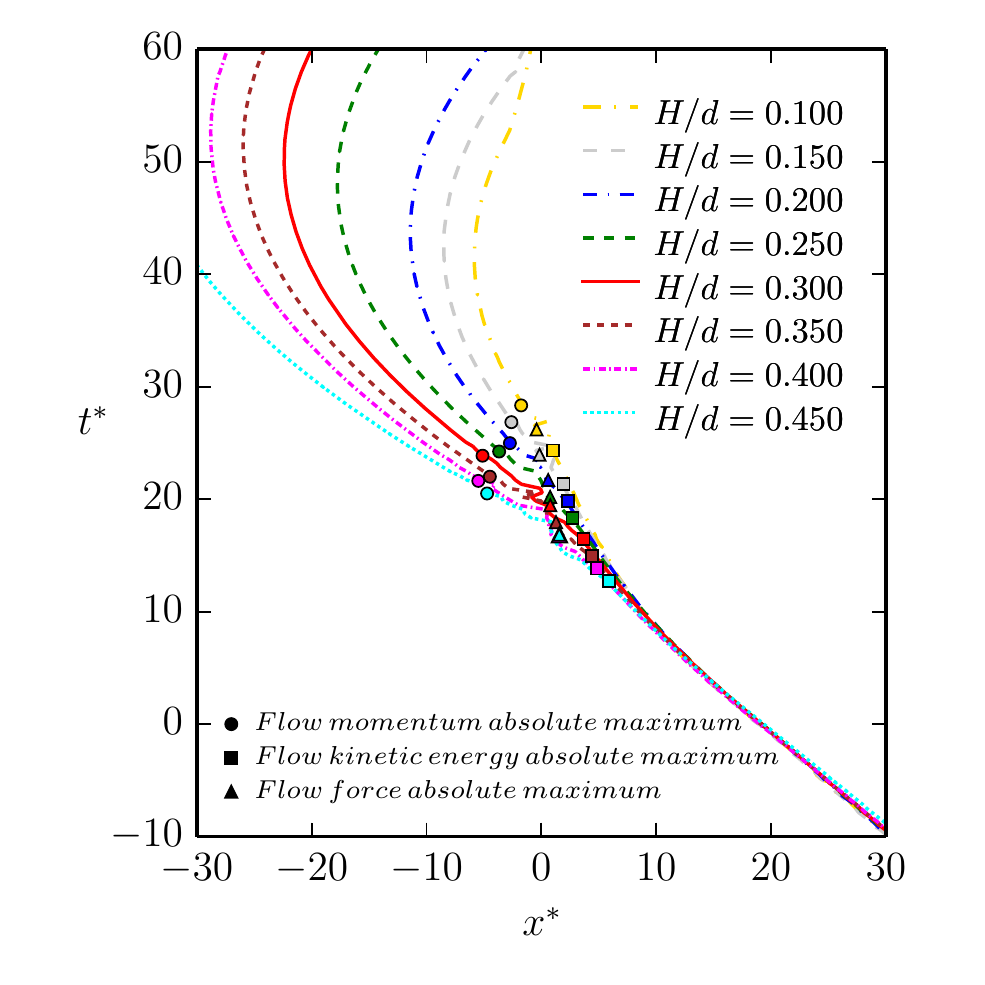} 
\caption{Wave front path of solitary waves of GPUSPH simulations for different ratios of $H/d$. $x^*=x/d$ and $t^*=t\cdot\sqrt{g/d}$. Origin of normalized time is set when wave front reaches the vertical of the beach toe, $X_o$, shown in Fig.~\ref{Scheme}. Squares, circles and triangles represent the absolute maxima of flow momentum, the absolute maxima of flow kinetic energy and the absolute maxima of flow force respectively.}
\label{curves}
\end{figure}

\begin{figure} 
\centering
\includegraphics[width=0.7\textwidth]{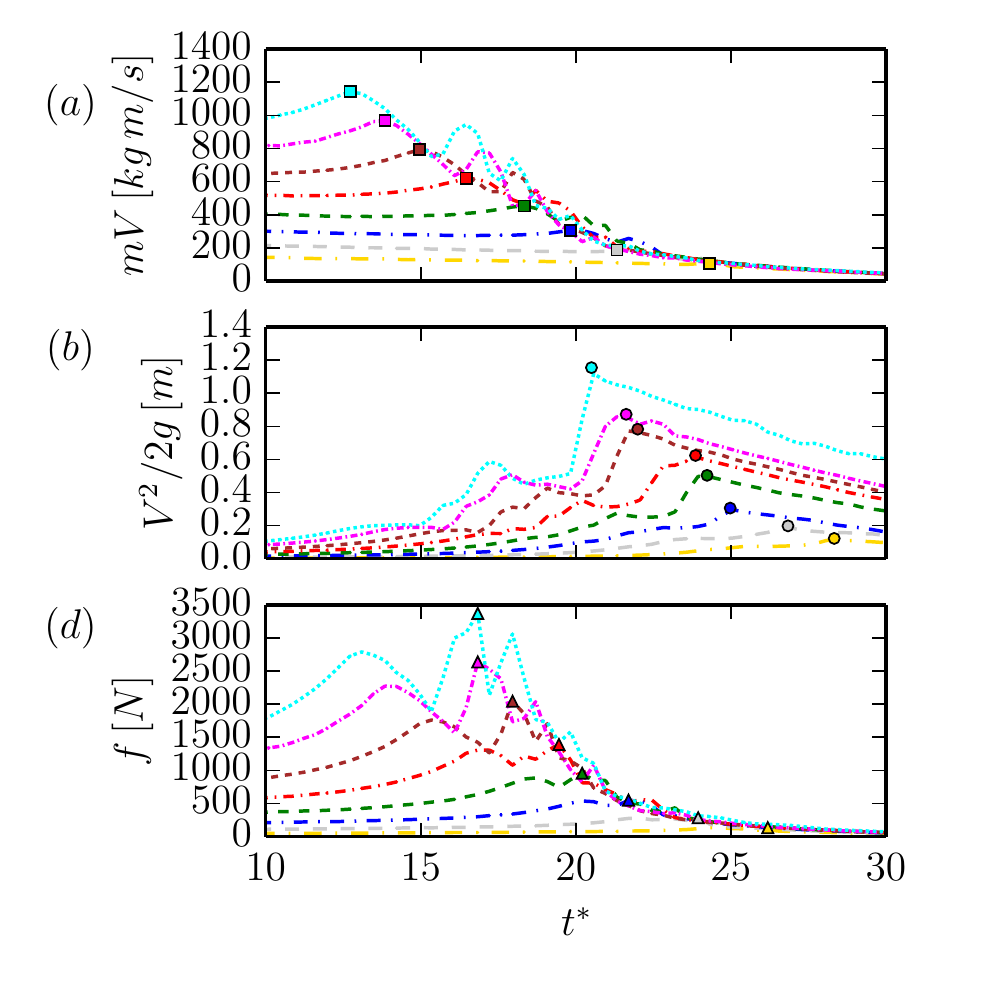} 
\caption{Subfigure $a)$ shows the time evolution of the maxima flow momentum for different $H/d$ ratios and its absolute maxima with squares. Subfigure $b)$ depicts the time evolution of the maxima flow kinetic energy for different $H/d$ ratios and its absolute maxima with circles. Subfigure $c)$ shows the time evolution of the maxima flow force for different $H/d$ ratios and its absolute maxima with triangles. For legend use the one shown in Fig. 5. Note that all values are per volume of fluid contained within the segments where they are computed from.}
\label{maxima}
\end{figure}

\begin{figure}
\centering
\includegraphics[width=0.7\textwidth]{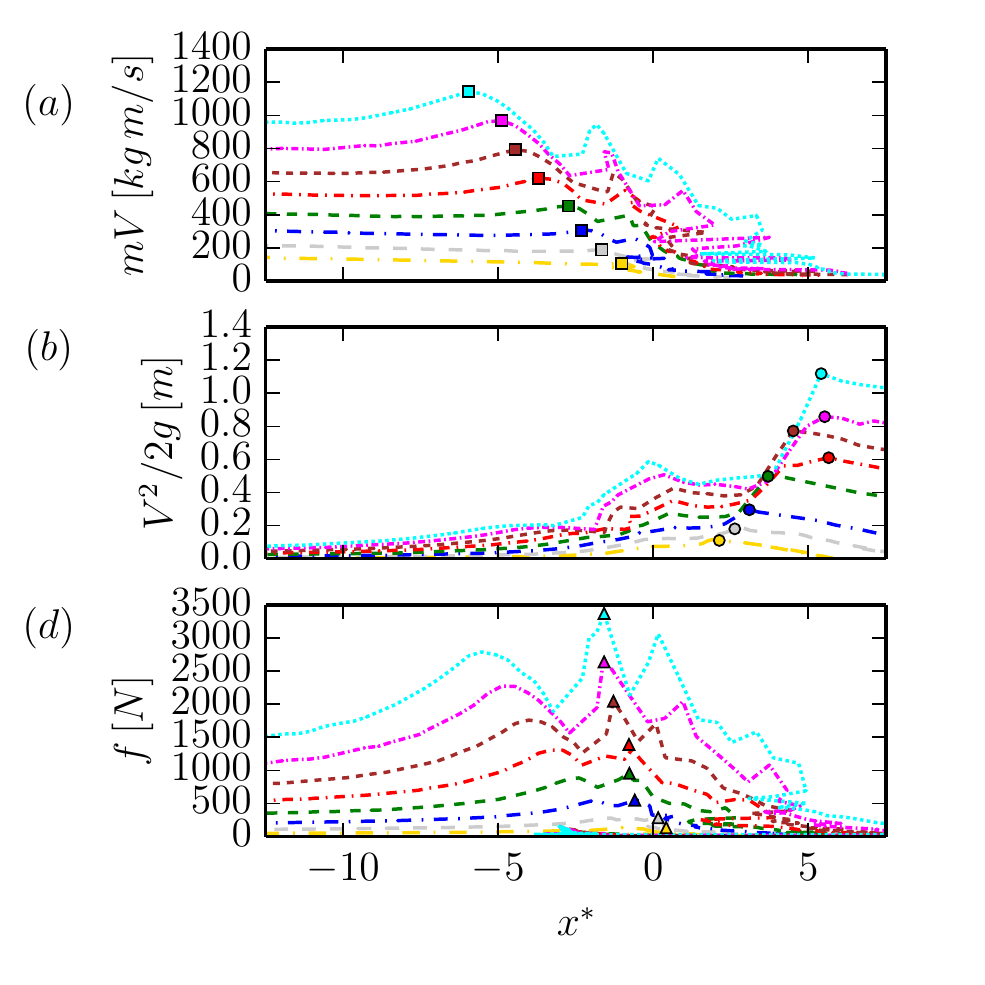} 
\caption{Subfigure $a)$ shows the spatial evolution of the maxima flow momentum for different $H/d$ ratios and its absolute maxima with squares. Subfigure $b)$ depicts the spatial evolution of the maxima kinetic energy for different $H/d$ ratios and its absolute maxima with circles. Subfigure $c)$ shows the spatial evolution of the maxima flow force for different $H/d$ ratios and its absolute maxima with triangles. For legend use the one shown in Fig. 5. Note that all values are per volume unit of fluid contained within the segments where they are computed from.}
\label{space}
\end{figure}

\begin{figure}
\centering
\includegraphics[width=0.35\textwidth]{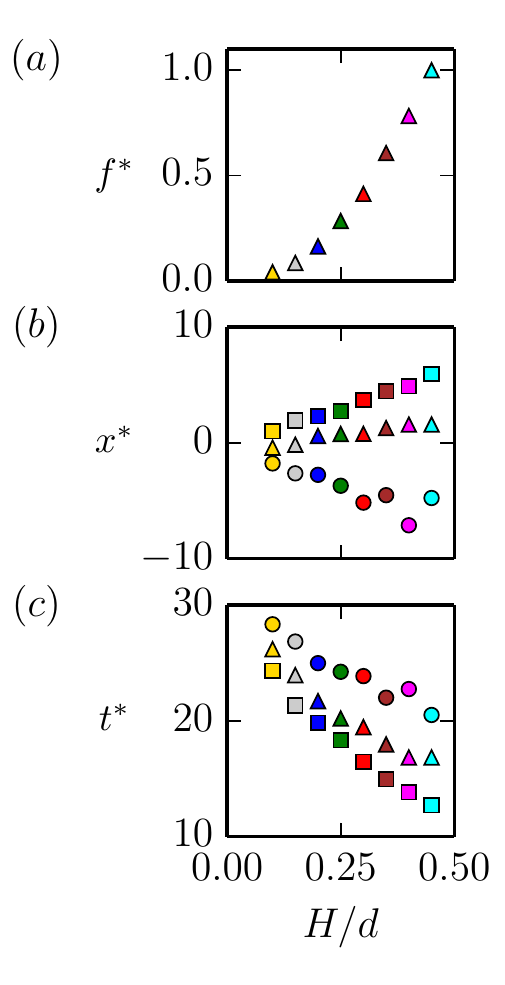} 
\caption{Subfigure $a)$ provides the relationship between the absolute maxima of the flow force normalized with the  value from the $H/d=0.45$ run with the $H/d$ ratio. Subfigure $b)$ shows the relationship between the location where the absolute maxima of the flow momentum, flow kinetic energy and flow force occur and the $H/d$ ratio. Subfigure $c)$ depicts the relationship between the time when they appear at with the $H/d$ ratio. As shown in Fig.~\ref{curves},~\ref{maxima} and~\ref{space}, squares, circles and triangles represent the absolute maxima of the flow momentum, absolute maxima of the flow kinetic energy and the absolute maxima of the flow force respectively for each $H/d$ simulated. For colors  use the legend shown in Fig. 5.}
\label{lines}
\end{figure}

Figure 3.5.5 in \cite{synolakis1986runup} presents the comparison of the wave front path, for a $H/d=0.0185$ solitary wave, between experiments and the solution of the nonlinear theory. He noted the presence of an intense wave front acceleration in the shoreline environment that has to be analyze. We study this effect by simulating solitary waves with different $H/d$ ratios and show the dimensionless spatial-temporal evolutions of their wave front paths in Fig.~\ref{curves} \cite[Fig. 3.5.5]{synolakis1986runup}. Note in Fig. 5 that the more horizontal the slope of the wave path is, higher its velocity is and vice versa. We observe a slope change of the wave front trajectory around the shoreline. Here is where flow accelerations occur and where we focus. The maxima of the flow momentum, flow kinetic energy and flow force are represented with different dashed lines in Fig.~\ref{curves},~\ref{maxima},~\ref{space} and~\ref{lines}. The format of these lines is consistent among these figures.The absolute maxima of flow momentum, flow kinetic energy and flow force is represented consistently with squares, circles and triangles respectively in these figures.

The temporal evolution of the maxima cited previously and their absolute maxima are presented in Fig.~\ref{maxima} for different $H/d$ ratios. Figure~\ref{maxima}a shows the evolutions of the maxima flow momentum in lines and their absolute maxima in squares. Figure~\ref{maxima}b depicts the maxima flow kinetic energy evolution and their absolute maxima with circles. The maximum kinetic energy evolution has two peaks. The first and smaller of the peaks occur due to the wave breaking. Figure~\ref{maxima}c shows the
maxima flow force evolution and their absolute maxima with triangles.

For the different $H/d$ solitary waves studied, Fig.~\ref{space} shows the spatial evolution of the variables mentioned in Fig.~\ref{maxima}. Figure~\ref{space}a shows the maxima flow momentum over space and their absolute maxima with squares and Fig.~\ref{space}b depicts the maxima flow kinetic energy over space and their absolute maxima with circles. As in Fig.~\ref{maxima}b, the maxima kinetic energy evolution has two peaks as well. Moreover Fig.~\ref{space}c shows the maxima flow force over space and its absolute maximum with triangles. Note that the absolute maxima of the flow force occurs just before the shoreline and where the maxima flow momentum reach their local maximum or second peaks. These locations are also where the maxima flow kinetic energy experience sharp increases.

We relate in Fig.~\ref{lines}a the relationship between the value of the flow force absolute maximum with $H/d$. Absolute maxima of flow force is normalized in this figure using the maximum value which is provided by the $H/d=0.45$ ratio simulation. Figure~\ref{lines}b shows the relationship between the locations of the absolute maxima of the flow momentum (squares), the absolute maxima of the flow kinetic energy (circles) and the absolute maxima of the flow force (triangles) with the $H/d$ ratio. Furthermore Fig.~\ref{lines}c displays the relationships between the times of the flow momentum absolute maxima (squares), flow kinetic energy absolute maxima (circles) and flow force absolute maxima (triangles) with the $H/d$ ratio.

			\section*{Discussion} 

In this contribution, we studied the three-dimensional breaking process of several solitary waves using GPUSPH model. Variations of the wave front velocity (accelerations) during breaking are analyzed in Fig.~\ref{curves}, where wave front velocities are represented by the inverse of the absolute values of the slopes of the curves. Note that the slopes of the curves are negative as the waves approaches the shoreline until the rundown starts. The more vertical the slope of the curve, the slower the wave front and vice versa. So, if the slope of the curve is less steep, then there is acceleration of the wave front. These  accelerations occur between the squares and circles (Fig. 5, 6, 7 and 8). The higher the $H/d$ ratio, the higher the intensity of the acceleration of the wave front.

The presence of  wave front acceleration during breaking is related to wave shoaling and has implications in the spatio-temporal evolution of the flow momentum, flow kinetic energy and flow force. While shoaling, as the wave becomes higher and steeper, the maximum flow momentum also increases. The maximum elevation of the wave crest generates the absolute maximum of the flow momentum (squares in Fig.~\ref{curves}, Fig.~\ref{maxima}a and Fig.~\ref{space}a). After this absolute maximum, the maximum flow momentum decreases because the wave becomes lower and the water depth  shallower.

For flow kinetic energy we observe that the largest velocities of the wave tip occur during breaking. However, the averaged velocity of the wave tip segment is not the largest because the averaging also considers slower water located deeper. The maximum flow kinetic energy reaches its absolute maximum when the water becomes shallower (circles in Fig.~\ref{maxima}b). It is important to note that circles in Fig.~\ref{maxima}b are also the locations where wave front and wave crest converge.

When the absolute maximum flow momentum occurs (squares in Fig.~\ref{maxima}a), the maximum kinetic energy at that time is very low compared with its absolute maximum (circles in Fig.~\ref{maxima}b).
Conversely, when the flow is at its absolute maximum kinetic energy (circles in Fig.~\ref{maxima}b), the maximum flow momentum at that time is very low compared with its absolute maximum (squares in Fig.~\ref{maxima}a). Additionally, the water depth is shallow where the absolute maxima of the flow kinetic energy occur. Hence, although the kinetic energy is involved in the dangerousness of the tsunamis, it is not the only one to be considered. The maximum flow momentum and especially the maximum flow force should also be taken into account.

By analyzing the evolutions of maximum flow momentum and the maximum kinetic energy, we observe that the second peak of the maximum momentum (Fig. 6a) almost coincides with the first peak of the maximum kinetic energy (Fig. 6b) and concurs with the sharp increment of the maximum kinetic energy (Fig. 7a and 7b). From hereon, the aim is to employ a variable that has an absolute maximum at points where the maximum flow momentum and maximum flow kinetic energy have their local maxima that is nearly coincident both spatially and temporally. Figures~\ref{maxima}c and \ref{space}c depicts the spatio-temporal evolution of the above mentioned variable which is given by the flow force. The triangles in both figures are points of absolute maximum of flow force which synchronizes with local maxima of flow momentum and flow kinetic energy.
The absolute maximum of the flow force appears just after the wave tip falls down on the water surface. This fall down causes a highly chaotic and turbulent layer of flow with significantly higher velocity at the surface and its vicinity. This process suddenly increases the kinetic energy (first sharp increase in Fig.~\ref{maxima}b and in Fig.~\ref{space}b) and the water level which leads to an increase in the flow momentum (local maxima peaks in Fig.~\ref{maxima}a and Fig.~\ref{space}a). Both together lead to the absolute maximum of the flow force (triangles in Fig.~\ref{maxima}c and Fig.~\ref{space}c).
This location and time is when the flow becomes the most turbulent and dangerous. Absolute maximum of flow force occurs just before shoreline (triangles around $x^*=0$ in Fig.~\ref{space}c.) regardless of the $H/d$ ratio of the wave. For the larger waves, $H/d$ = 0.45, 0.40 and 0.35 runs, secondary peaks are present following the absolute maxima of the flow force, adding extra hazards. These results have to be considered in the shoreline facilities design.

Figure~\ref{lines}a depicts the absolute maxima of the flow force as a function of $H/d$. This figure allows to estimate the absolute maximum flow force for any solitary wave not studied in this work with the same setup.
Additionally, Fig.~\ref{lines}b and \ref{lines}c provide the location and time respectively of the absolute maximum of flow momentum (squares), the absolute maximum of flow kinetic energy (circles) and the absolute maximum of flow force (triangles). The larger the $H/d$ ratio is, the larger the distance between squares and circles gets (see Fig.~\ref{lines}b and Fig.~\ref{lines}c) and also the longer the breaking process takes and the farther the runup reaches. Therefore by knowing the $H/d$ ratio of any solitary wave, it is possible to estimate not only the magnitude of the flow force, but also its occurrence spatially and temporally.

			\section*{Conclusions}

In this study, we carried out numerical simulations with a non-depth integrated formulation of solitary waves to study their hazards during their proximity to the shoreline. Because wave breaking is a three-dimensional process, neither the shallow water equations nor the Boussinesq approximations can describe breaking analysis accurately. We approached this by averaging momentum, kinetic energy and force of the flow in segments along $x$ axis and by obtaining their maxima evolutions from three-dimensional SPH simulations. Figure 5 shows the wave front paths for different $H/d$ ratios. From Fig. 6 and 7 we obtain the locations and time of the absolute maxima of flow momentum, flow kinetic energy and flow force for various $H/d$. When these points are plotted on Fig. 5 they appear on the curve of their respective $H/d$ wave path. This implies that the absolute maximum of flow momentum, the absolute maximum of flow kinetic energy and the absolute maximum of flow force always occur in the wave front, regardless of the $H/d$ ratio. So we conclude that the wave front will always be the most dangerous part of the tsunami at any time regardless of the $H/d$ wave ratio.

Also we conclude that flow force is the most important variable to utilize, over the flow momentum or flow kinetic energy, in order to identify the tsunamis dangerousness during breaking. This is mainly because the peaks of the absolute maxima of the flow force occur at points of nearly simultaneous local maxima of flow momentum and flow kinetic energy. This also occurs regardless of their $H/d$ ratio. The absolute maximum flow force for any $H/d$ occurs just before the shoreline. It denotes that the areas where the absolute maximum flow force occurs are most prone to hazards during the breaking and inundation processes.

These conclusions render a better understanding of the breaking process and let us identify the physical variables that must be considered to evaluate the destructive hazards of solitary waves in space and time. Also they provide important considerations to be taken into account to develop more reliable tsunami-risk evaluations in order to protect marine infrastructures and human lives.

			\section*{Acknowledgements}

The work presented in here is based upon work partially supported by the National Science Foundation under grants NSF-CMMI-1208147 and NSF-CMMI-1206271.

\section*{References}

\bibliographystyle{elsarticle-harv}


\end{document}